\documentclass[10pt,conference]{IEEEtran}\IEEEoverridecommandlockouts
\usepackage{ epsfig,graphics,subfigure,psfrag,amsmath,cases}
\usepackage{latexsym,amssymb,amsmath,epsfig,subfigure,algorithm,amsthm}
\usepackage{algorithmic}
\usepackage{color}
\usepackage{url}
\usepackage{scrtime}
\usepackage{bm}
\usepackage{graphicx}

\usepackage{pbox}
\usepackage{framed}
\usepackage{color}
\usepackage[makeroom]{cancel}

\author{Elena Boshkovska,
 Nikola Zlatanov, Linglong Dai,   Derrick Wing Kwan Ng, and Robert Schober\thanks{ E. Boshkovska and R. Schober are with Friedrich-Alexander-University Erlangen-N\"urnberg (FAU), Germany.  Linglong Dai is with Tsinghua University, Beijing, China.  L. Dai is supported by the International Science \& Technology Cooperation Program of China (Grant No. 2015DFG12760) and the National Natural Science Foundation of China (Grant No. 61571270).
D. W. K. Ng is with The University of New South Wales, Australia. N. Zlatanov is with Monash University, Australia. R. Schober is supported by the AvH Professorship Program of the Alexander von Humboldt Foundation. D. W. K. Ng is supported under Australian Research Council's Scheme Discovery Early Career Researcher Award funding scheme (project
number DE170100137).}

}

\title{Secure SWIPT Networks Based on a  Non-linear Energy Harvesting Model }

\date{\thistime,\,\today}

\newtheorem{Thm}{Theorem}

\newtheorem{Prob}{Problem}
\newtheorem{proposition}{Proposition}

\newcommand{\abs}[1]{\lvert#1\rvert}

\newcommand{\norm}[1]{\lVert#1\rVert}
\DeclareMathOperator{\Tr}{\mathrm{Tr}}
\DeclareMathOperator{\zero}{\mathbf{0}}
\DeclareMathOperator{\Rank}{\mathrm{Rank}}

\DeclareMathOperator{\maxo}{\mathrm{maximize}}
\DeclareMathOperator{\mino}{\mathrm{minimize}}

\begin{document}
\IEEEspecialpapernotice{(Invited Paper)}
\maketitle

\begin{abstract}
We optimize resource allocation to enable communication security  in simultaneous wireless information and power transfer (SWIPT) for internet-of-things (IoT) networks. The resource allocation algorithm design is formulated as a non-convex optimization problem. We aim at maximizing the total harvested power at energy harvesting (EH) receivers via the joint optimization of transmit beamforming vectors and the covariance matrix of the artificial noise injected to facilitate secrecy provisioning. The proposed problem formulation takes into account the non-linearity of energy harvesting circuits and the   quality of service requirements for secure communication.
  To obtain a globally optimal solution of the resource allocation problem, we first transform the resulting non-convex sum-of-ratios objective function into an equivalent objective function in parametric subtractive form, which facilitates the design of a novel iterative resource allocation algorithm. In each iteration, the semidefinite programming (SDP)  relaxation approach is adopted to solve a rank-constrained optimization problem optimally. Numerical results reveal that the proposed algorithm can guarantee communication security and provide a significant performance gain in terms of the harvested energy  compared to existing designs which are based on  the traditional linear EH model.
\end{abstract}

\renewcommand{\baselinestretch}{0.91}
\large\normalsize

\section{Introduction}

In the era of the Internet of Things (IoT), it is expected that $50$ billion wireless communication devices will be connected worldwide \cite{JR:IOT}. Smart physical objects equipped with sensors and wireless communication chips are able to collect and exchange information. These smart objects are wirelessly connected to computing systems to provide intelligent everyday services  such as e-health, automated control, energy management (smart city and smart grid), logistics, security control, and safety management, etc. However,  the  limited energy storage capacity of battery-powered wireless communication devices severely limits the lifetime of wireless communication networks. Although battery replacement provides an intermediate solution to energy shortage, frequent replacement of batteries can be costly and cumbersome. This creates a serious performance bottleneck in providing stable  communication services. On the other hand, a promising approach to extend the lifetime of wireless communication networks is to equip wireless communication devices with energy harvesting (EH) technology to scavenge energy from  external sources. Solar, wind, tidal, biomass, and geothermal are the major  renewable energy sources for generating electricity \cite{JR:Kwan_hybrid_BS}. Yet, these conventional natural energy sources are usually climate and location dependent which limits the mobility of the wireless devices. More importantly, the intermittent and uncontrollable nature of these natural energy sources is a major obstacle   for providing stable wireless communications via traditional EH technologies.

Recently, wireless energy transfer (WET) has attracted significant attention
from both academia and industry \cite{CN:Shannon_meets_tesla}\nocite{Krikidis2014,Ding2014,JR:SWIPT_mag,JR:SWIPT_mag_Ming_Kwan, JR:QQ_WPC,COML:EE_WIPT,JR:MIMO_WIPT}--\cite{JR:QQ_WPCN}, as a key to unlock the potential of IoT. Generally speaking, the  WET technology can be divided into three categories: magnetic resonant coupling, inductive coupling,  and radio frequency (RF)-based WET. The first two technologies rely on near-field magnetic fields and do not support any mobility of EH devices, due to the short wireless charging distances and the required alignment of the magnetic field with the EH circuits. In contrast, RF-based WET technologies \cite{CN:Shannon_meets_tesla}\nocite{Krikidis2014,Ding2014,JR:SWIPT_mag,JR:SWIPT_mag_Ming_Kwan, JR:QQ_WPC,COML:EE_WIPT,JR:MIMO_WIPT}--\cite{JR:QQ_WPCN} utilize the far-field of electromagnetic (EM) waves which enables concurrent
wireless charging and data communication in WET networks over long distances (e.g. hundreds of metres).  Moreover, the broadcast nature of wireless channels facilitates one-to-many wireless charging  which eliminates the need for power cords and manual recharging for IoT devices. As a result,  simultaneous wireless information and power transfer (SWIPT) is expected to be a key enabler for sustainable IoT communication networks. Yet, the introduction of SWIPT to communication systems has led to a paradigm shift in both system architecture and resource allocation
algorithm design. For instance, in SWIPT systems, one can increase the energy
of the information carrying signal to increase the amount of   RF energy harvested at the receivers. However,  increasing the power of the information signals may also increase their susceptibility to
eavesdropping, due to the higher potential for information leakage. As a result, both communication security concerns and the need for efficient WET  naturally arise in systems providing SWIPT services.

 Nowadays, various types of cryptographic encryption algorithms are employed at the application layer for guaranteeing wireless communication security.  However,  secure secret key management and distribution via an authenticated third party is typically required for these algorithms, which may not be realizable in future wireless IoT networks due to the expected massive numbers of devices. Therefore, a considerable amount of work has recently been
devoted to information-theoretic physical (PHY) layer security
as a complementary technology to the existing encryption algorithms \cite{Report:Wire_tap}\nocite{JR:Massive_MIMO,JR:Artifical_Noise1,JR:Chen_security,JR:HM_security_1,JR:HM_security_2}--\cite{JR:Kwan_secure_imperfect}. It has been shown that
in a wire-tap channel, if the source-destination channel enjoys better conditions compared to the source-eavesdropper channel  \cite{Report:Wire_tap}, perfectly secure
  communication  between a source and a destination is possible. Hence,  multiple-antenna technology and advanced signal processing algorithms have been proposed to ensure secure communications. Specifically, by exploiting the extra degrees of freedom
offered by multiple antennas,  the information beams can be focused on the desired legitimate receivers to reduce the chance of information leakage. Besides, artificial noise can be injected  into the communication channel deliberately to degrade the channel quality of eavesdroppers. These concepts have also been extended to SWIPT systems to provide secure communication. In \cite{JR:Kwan_secure_imperfect}, beamforming was studied to enhance security and power efficiency in SWIPT systems.  The authors of \cite{JR:MOOP_SWIPT} proposed a multi-objective optimization framework to investigate the non-trivial tradeoff between interference, total harvested power, and energy consumption in a secure cognitive radio SWIPT network.   However, the resource allocation algorithms designed for secure SWIPT systems \cite{JR:Kwan_secure_imperfect,JR:MOOP_SWIPT} were based on a linear EH model which does not capture the highly non-linear characteristics of practical end-to-end WET \cite{CN:EH_measurement_2}\nocite{JR:Energy_harvesting_circuit,JR:EH_measurement_1}--\cite{JR:non_linear_model}.  In particular, existing resource allocation schemes  designed for the
 linear EH model may lead to severe
resource allocation mismatches resulting in performance
degradation in WET and secure communications. These observations motivate us to study the design of efficient  resource allocation algorithms for secure SWIPT systems taking into account a practical non-linear EH model.

 \begin{figure}[t]
 \centering
\includegraphics[width=3.5 in]{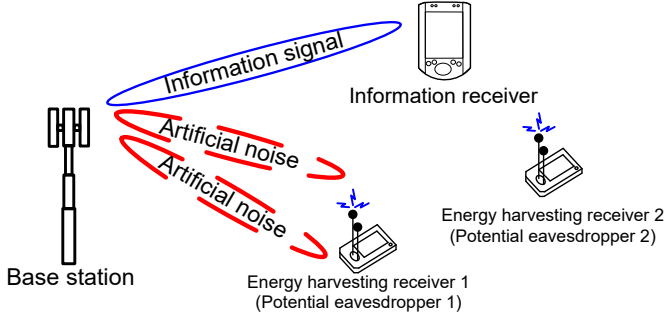}
 \caption{A  SWIPT model with an active IR and $J = 2$ ERs.
The ERs harvest wireless energy from the RF signal and are treated as
potential eavesdroppers by the base station. } \label{fig:system_model}
\end{figure}

\section{System Model}
In  this  section,  we first introduce the  notation  adopted  in  this
paper. Then,   we  present  the  downlink  channel
model  for  secure  communication in SWIPT systems.

\subsection{Notation} We use boldface capital and lower case letters to denote matrices and vectors, respectively. $\mathbf{A}^H$, $\Tr(\mathbf{A})$, $\Rank(\mathbf{A})$, and $\det(\mathbf{A})$ represent the Hermitian transpose, trace, rank, and determinant of  matrix $\mathbf{A}$, respectively;  $\mathbf{A}\succ \zero$ and $\mathbf{A}\succeq \zero$ indicate that $\mathbf{A}$ is a positive definite and a  positive semidefinite matrix, respectively; $\mathbf{I}_N$ is the $N\times N$ identity matrix; $[\mathbf{q}]_{m:n}$  returns a vector with the $m$-th to the $n$-th elements of vector $\mathbf{q}$; $\mathbb{C}^{N\times M}$ denotes the set of all $N\times M$ matrices with complex entries; $\mathbb{H}^N$ denotes the set of all $N\times N$ Hermitian matrices.  The circularly symmetric complex Gaussian (CSCG) distribution is denoted by ${\cal CN}(\mathbf{m},\mathbf{\Sigma})$ with mean vector $\mathbf{m}$ and covariance matrix $\mathbf{\Sigma}$; $\sim$ indicates ``distributed as"; ${\cal E}\{\cdot\}$ denotes  statistical expectation; $\abs{\cdot}$ represents the absolute value of a complex scalar. $[x]^+$ stands for $\max\{0,x\}$ and   $[\cdot]^T$ represents the  transpose
operation.

\subsection{Channel Model}\label{sect:model}
A frequency flat fading channel for downlink communication is considered. The SWIPT system comprises
 a base station (BS), an  information receiver (IR),
 and $J$ energy harvesting receivers (ER), as shown in  Figure \ref{fig:system_model}. The  BS   is equipped with $N_{\mathrm{T}}\geq 1$ antennas.  The IR  is a single-antenna device  and each ER is equipped with $N_{\mathrm{R}}\geq 1$ receive antennas for EH.  In the considered system, the signal intended for the IR is overheard by the ERs due to the broadcast nature of wireless channels. To guarantee communication security,  the ERs are treated as potential eavesdroppers which has to be taken into account for resource allocation algorithm design. We assume that  $N_{\mathrm{T}}> N_{\mathrm{R}}$
 for the following study. The signals received  at the IR  and  ER  $j\in\{1,\ldots, J\}$ are modelled as
\begin{eqnarray}
y&=&\mathbf{h}^H(\mathbf{w}s+\mathbf{v}) +n,\,\, \mbox{and}\\
\mathbf{y}_{\mathrm{ER}_j}&=&\mathbf{G}_j^H(\mathbf{w}s+\mathbf{v})+\mathbf{n}_{\mathrm{ER}_j},\,\,  \forall j\in\{1,\dots,J\},
\end{eqnarray}respectively, where $s\in\mathbb{C}$ and $\mathbf{w}\in\mathbb{C}^{N_{\mathrm{T}}\times1}$  are the information symbol  and  the corresponding beamforming vector, respectively.  Without loss of generality, we assume that ${\cal E}\{\abs{s}^2\}=1$.  $\mathbf{v}\in\mathbb{C}^{N_{\mathrm{T}}\times 1}$ is an artificial noise vector generated by the BS to facilitate efficient WET and to guarantee communication security.  In particular,  $\mathbf{v}$ is modeled as a random vector with circularly
symmetric complex Gaussian distribution
\begin{eqnarray}
\mathbf{v}\sim {\cal CN}(\mathbf{0}, \mathbf{V}),
\end{eqnarray}
where $\mathbf{V}\in \mathbb{H}^{N_{\mathrm{T}}}, \mathbf{V}\succeq \mathbf{0}$, denotes the covariance matrix of the artificial noise.  The channel vector between the  BS  and the IR  is denoted by $\mathbf{h}\in\mathbb{C}^{N_{\mathrm{T}}\times1}$ and the channel matrix between the  BS  and  ER  $j$  is denoted by $\mathbf{G}_j\in\mathbb{C}^{N_{\mathrm{T}}\times N_{\mathrm{R}}}$. $n\sim{\cal CN}(0,\sigma_{\mathrm{s}}^2)$ and $\mathbf{n}_{\mathrm{ER}_j}\sim{\cal CN}(\zero,\sigma_{\mathrm{s}}^2\mathbf{I}_{N_{\mathrm{R}}})$ are the additive white Gaussian noises (AWGN) at  the  IR  and ER  $j$, respectively, where $\sigma_{\mathrm{s}}^2$ denotes the noise power at each antenna of the receiver.


\subsection{Energy Harvesting Model}
\begin{figure}[t]
 \centering
\includegraphics[width=3.5 in]{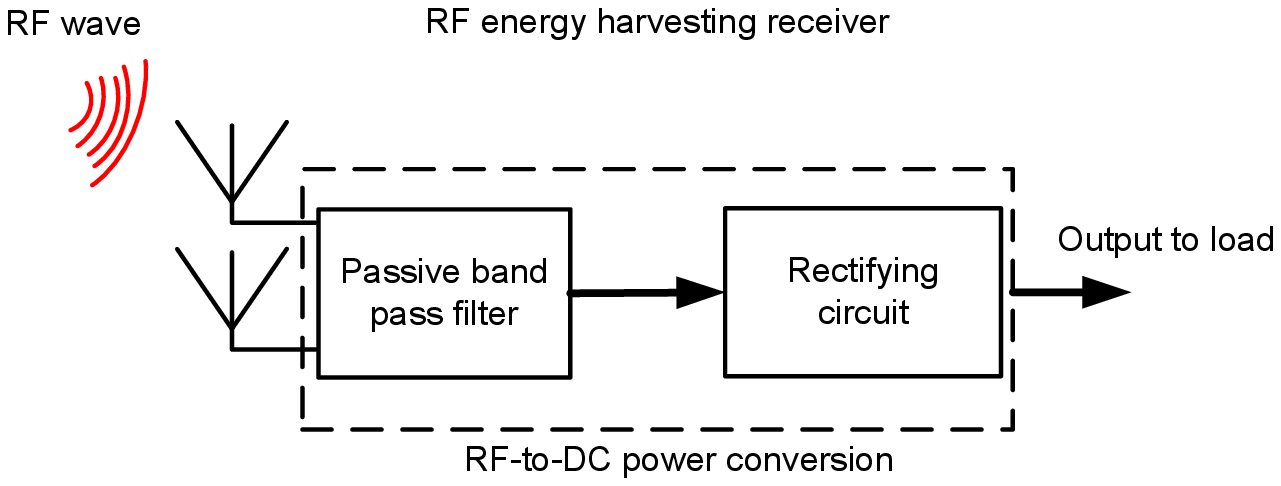}
 \caption{Block diagram of an ER. } \label{fig:circuit_model}
 \end{figure}

\begin{figure}[t]
 \centering

\includegraphics[width=3.5 in]{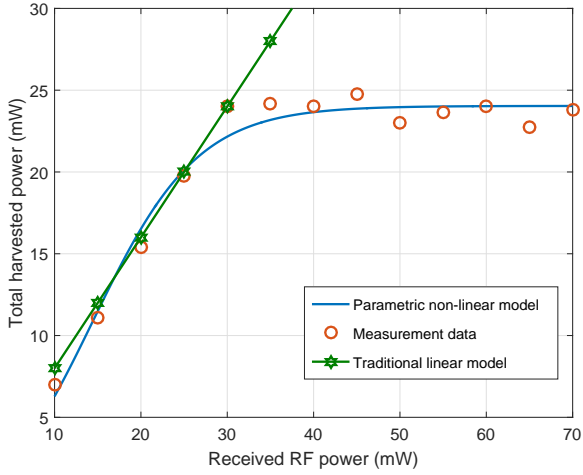}
 \caption{A comparison between experimental data from \cite{CN:EH_measurement_2}, the harvested power for the non-linear model in \eqref{eqn:EH_non_linear}, and the linear EH model with $\eta_j=0.8$ in \eqref{eqn:linear_EH_model}. } \label{fig:comparsion_EH}
\end{figure}

Figure \ref{fig:circuit_model} depicts the block diagram of the ER in SWIPT systems. In general, a bandpass filter and a
rectifying circuit  are adopted in an RF-ER to convert the received RF power to direct current (DC) power.  The total received RF power at ER  $j$  is given by \begin{eqnarray}
P_{\mathrm{ER}_j}=\Tr\Big((\mathbf{w}\mathbf{w}^H +\mathbf{V})\mathbf{G}_j\mathbf{G}_j^H\Big).
\end{eqnarray}
In the SWIPT literature, for simplicity, the total harvested power at  ER  $j$ is typically modelled as follows:
\begin{eqnarray}\label{eqn:linear_EH_model}
\Phi_{\mathrm{ER}_j}^{\mathrm{Linear}}=\eta_j P_{\mathrm{ER}_j},
\end{eqnarray}
 \noindent
where $0\leq\eta_j\leq1$ denotes the energy conversion efficiency of ER $j$. From \eqref{eqn:linear_EH_model}, it can be seen that with existing models,  the total harvested power at the ER is linearly and directly proportional to the received RF power. However, practical RF-based EH circuits  consist of
resistors, capacitors, and diodes. Experimental results have shown that  these circuits  \cite{CN:EH_measurement_2}\nocite{JR:Energy_harvesting_circuit}--\cite{JR:EH_measurement_1} introduce various non-linearities into the end-to-end WET.  In order to design a resource allocation algorithm for practical secure SWIPT systems,  we adopt the non-linear parametric EH model from \cite{JR:non_linear_model,JR:Elena_TCOM}.  Consequently, the total harvested power at ER  $j$, $\Phi_{\mathrm{ER}_j}$, is modelled as:
 \begin{eqnarray}\label{eqn:EH_non_linear}
\hspace*{-3mm}\Phi_{\mathrm{ER}_j}\hspace*{-1.0mm}&=&\hspace*{-1.0mm}
 \frac{[\Psi_{\mathrm{ER}_j}
 - M_j\Omega_j]}{1-\Omega_j},\, \Omega_j=\frac{1}{1+\exp(a_jb_j)},\\
\mbox{where}\,\,\Psi_{\mathrm{ER}_j}\hspace*{-1.0mm}&=&\hspace*{-1.0mm} \frac{M_j}{1+\exp\Big(-a_j(P_{\mathrm{ER}_j}-b_j)\Big)}
  \end{eqnarray}
 is a sigmoid function which takes the received RF power, $P_{\mathrm{ER}_j}$,  as the input.
  Constant $M_j$ denotes the maximal harvested power at ER  $j$  when the EH circuit is driven to  saturation due to  an exceedingly large input RF power.    Constants $a_j$ and $b_j$  capture the joint effects of  resistance, capacitance, and circuit sensitivity. In particular,  $a_j$ reflects the non-linear charging rate (e.g. steepness of the curve) with respect to the input power and $b_j$ determines to the minimum turn-on voltage of the EH circuit.
  In practice,  parameters $a_j$, $b_j$, and $M_j$ of the proposed model in \eqref{eqn:EH_non_linear}
can be obtained using a standard curve fitting algorithm for measurement results of a given EH hardware circuit.  In Figure~\ref{fig:comparsion_EH}, we show an example for the curve fitting for the non-linear EH model in \eqref{eqn:EH_non_linear} with parameters $M=0.024$, $b=0.014$, and $a=150$. It can be observed that the parametric non-linear model closely matches experimental results provided in \cite{CN:EH_measurement_2} for the wireless power harvested by a practical EH circuit. Figure~\ref{fig:comparsion_EH} also illustrates the inability of the linear model in \eqref{eqn:linear_EH_model} to capture the non-linear characteristics of practical EH circuits, especially in the high received RF power regime.
  \noindent

\subsection{Secrecy Rate}
 Assuming perfect channel state information (CSI) is available at the
receiver for coherent detection, the  achievable rate (bit/s/Hz) between the  BS  and the IR is given by
\begin{eqnarray}
R&=&\log_2\Big(1+\frac{\mathbf{w}^H\mathbf{H}\mathbf{w}}
{\Tr(\mathbf{H}\mathbf{V})+\sigma_{\mathrm{s}}^2}\Big),
\end{eqnarray}
where $\mathbf{H}=\mathbf{h}\mathbf{h}^H$.

On the other hand, the capacity between the  BS  and ER $j$ for decoding the signal of the IR  can be expressed as
\begin{eqnarray}\label{eqn:Capacity_eve}
R_{\mathrm{ER}_j}&=&\log_2\det(\mathbf{I}_{N_{\mathrm{R}}}+
\mathbf{Q}_{j}^{-1}
\mathbf{G}_j^H\mathbf{w}\mathbf{w}^H\mathbf{G}_j),\\
\notag\mathbf{Q}_{j}&=&\mathbf{G}_j^H\mathbf{V}\mathbf{G}_j+
\sigma_{\mathrm{s}}^2\mathbf{I}_{N_{\mathrm{R}}}\succ \zero,
\end{eqnarray}
where $\mathbf{Q}_{j}$ denotes the  interference-plus-noise covariance matrix for ER $j$. Hence, the  achievable secrecy rate of the IR is given by \cite{JR:Artifical_Noise1}
\begin{eqnarray}\label{eqn:secure_cap}
R_{\mathrm{sec}}&=&\Big[R-\underset{\forall j}{\max}\,\{ R_{\mathrm{ER}_j}\}\Big]^+.
\end{eqnarray}

\section{Optimization Problem and Solution}\label{sect:problem_formulation_solution}
In the considered SWIPT system, we aim to maximize the total harvested power in the system  while providing secure communication to the IR. To this end, we formulate the resource allocation
algorithm design as the following non-convex optimization
problem assuming that perfect channel state information is available \footnote{In the sequel,  since $\Omega_j$ does not affect the design of the optimal resource allocation policy, with a slight abuse of notation,  we will directly use $\Psi_{\mathrm{ER}_j}$ to represent the harvested power at ER $j$ for simplicity  of presentation. }:
\begin{Prob}{ \normalfont Resource Allocation for Secure SWIPT:}
\begin{eqnarray}\label{eqn:TP_maximization}
\underset{\mathbf{V}\in \mathbb{H}^{N_{\mathrm{T}}},\mathbf{w}}{\maxo}\,\, && \sum_{j=1}^J \Psi_{\mathrm{ER}_j}\\
\mathrm{subject\,\,to}\,\, &&\mathrm{C1}:\,\,\norm{\mathbf{w}}_2^2 + \Tr(\mathbf{V})\leq P_{\mathrm{max}},\notag\\
&&\mathrm{C2}:\,\, \frac{\mathbf{w}^H\mathbf{H}\mathbf{w}}
{ \Tr(\mathbf{V}\mathbf{H})+\sigma_{\mathrm{s}}^2} \geq \Gamma_{\mathrm{req}},\notag\\
&&\mathrm{C3: }R_{\mathrm{ER}_j}\le R_{\mathrm{ER}}^{\mathrm{Tol}},\,\, \forall j,\notag\\
&& \mathrm{C4}:\,\, \mathbf{V}\succeq \zero.\notag
\end{eqnarray}\end{Prob}\vspace*{-3mm}
\noindent Constants $P_{\max}$ and $\Gamma_{\mathrm{req}}$  in constraints C1 and C2 denote the maximum transmit power budget and the minimum required signal-to-interference-plus-noise ratio (SINR) at the IR, respectively.  Constant $R_{\mathrm{ER}}^{\mathrm{Tol}}>0$ in C3 is the maximum tolerable data rate which restricts the capacity of ER $j$ if it  attempts to decode the signal intended for the IR.    In practice, the  BS  sets $\log_2(1+\Gamma_{\mathrm{req}})> R_{\mathrm{ER}}^{\mathrm{Tol}}>0$, to ensure secure communication\footnote{ In the considered problem formulation, we can  guarantee  that the achievable secrecy rate is bounded below by $R_{\mathrm{sec}}\ge \log_2(1+\Gamma_{\mathrm{req}})-R_{\mathrm{ER}}^{\mathrm{Tol}}>0$ if the problem is feasible.}. Constraint C4 and $\mathbf{V}\in \mathbb{H}^{N_{\mathrm{T}}}$ constrain matrix $\mathbf{V}$  to be a positive semidefinite Hermitian matrix. It can be observed that the objective function in \eqref{eqn:TP_maximization} is a non-convex function due to its sum-of-ratios form. Besides, the log-det function in C3 is non-convex. Now, we first transform the non-convex objective function into an equivalent objective function in subtractive form via the following theorem.


\begin{Thm}\label{Thm:1}
Suppose $\{\mathbf{w}^*,\mathbf{V}^*\}$ is the optimal solution to   \eqref{eqn:TP_maximization}, then there exist two column vectors ${\bm \mu}^*=[\mu_1^*,\ldots,\mu_J^*]^T$ and ${\bm \beta}^*=[\beta_1^*,\ldots,\beta_J^*]^T$ such that $\{\mathbf{w}^*,\mathbf{V}^*\}$ is an optimal solution to the following optimization problem
\begin{equation} \label{eqn:transformed}
\underset{\mathbf{V}^*\in \mathbb{H}^{N_{\mathrm{T}}},\mathbf{w}^*\in {\cal F}}\maxo\,\sum_{j=1}^J \mu_j^*\Big[\hspace*{-0.5mm}M_j\hspace*{-0.5mm}- \hspace*{-0.5mm} \beta_j^*\Big(1+\exp\big(\hspace*{-0.5mm}-\hspace*{-0.5mm}a_j(P_{\mathrm{ER}_j}\hspace*{-0.5mm}-\hspace*{-0.5mm}b_j)\big)\Big)\hspace*{-0.5mm}\Big],
\end{equation}
where $\cal F$ is the feasible solution set of \eqref{eqn:TP_maximization}. Besides, $\{\mathbf{w}^*,\mathbf{V}^*\}$  also satisfies the following system of equations:
\begin{eqnarray} \label{eqn:conditions1}
\beta_j^*\Big(1+\exp\big(\hspace*{-0.5mm}-\hspace*{-0.5mm}a_j(P_{\mathrm{ER}_j}^*\hspace*{-0.5mm}-\hspace*{-0.5mm}b_j)\big)\Big)-M_j&=&0,\\
\mu_j^*\Big(1+\exp\big(\hspace*{-0.5mm}-\hspace*{-0.5mm}a_j(P_{\mathrm{ER}_j}^*\hspace*{-0.5mm}-\hspace*{-0.5mm}b_j)\big)\Big)-1&=&0, \label{eqn:conditions2}
\end{eqnarray}
and $P_{\mathrm{ER}_j}^*=\Tr\Big((\mathbf{w}^*(\mathbf{w^*})^H +\mathbf{V}^*)\mathbf{G}_j\mathbf{G}_j^H\Big)$.
\end{Thm}

\,\,\emph{Proof:} Please refer to \cite{JR:sum_of_ratios} for a proof of  Theorem 1. \qed

Therefore,  for  \eqref{eqn:TP_maximization},
we have an equivalent  optimization problem in \eqref{eqn:transformed} with an objective function in subtractive form with extra parameters $(\bm{\mu}^*,\bm{\beta}^*)$. More importantly, the two problems have the same optimal solution $\{\mathbf{w}^*,\mathbf{V}^*\}$.
Besides,  the optimization problem in \eqref{eqn:transformed} can be solved by an iterative algorithm  consisting of two nested loops \cite{JR:sum_of_ratios}.   In the inner loop, the optimization problem in \eqref{eqn:transformed} for given $(\bm{\mu},\bm{\beta})$, ${\bm \mu}=[\mu_1,\ldots,\mu_J]^T$ and ${\bm \beta}=[\beta_1,\ldots,\beta_J]^T$,  is  solved.  Then, in the outer loop,  we find the optimal $(\bm{\mu}^*,\bm{\beta}^*)$ satisfying the system of equations in \eqref{eqn:conditions1} and \eqref{eqn:conditions2}, cf. Algorithm 1 in Table \ref{table:algorithm}.
 \subsection{Solution of the Inner Loop Problem}
In each iteration, in line 3 of Algorithm 1,  we solve an inner loop optimization problem in its hypograph form:
\begin{Prob}{Inner Loop Problem}
\begin{eqnarray}\label{eqn:TP_maximization-transfomred}\notag
\underset{\mathbf{W},\mathbf{V}\in \mathbb{H}^{N_{\mathrm{T}}},\bm\tau}{\maxo}\,\, &&\sum_{j=1}^J \mu_j^*\Big[\hspace*{-0.5mm}M_j\hspace*{-0.5mm}- \hspace*{-0.5mm} \beta_j^*\Big(1+\exp\big(\hspace*{-0.5mm}-\hspace*{-0.5mm}a_j(\tau_j\hspace*{-0.5mm}-\hspace*{-0.5mm}b_j)\big)\Big)\hspace*{-0.5mm}\Big]\\
\hspace*{-1mm}\mathrm{subject\,\,to}\,\, &&\hspace*{-2mm}\mathrm{C1}:\,\,\Tr(\mathbf{W+V})\leq P_{\mathrm{max}},\notag\\
\hspace*{-1mm}&&\hspace*{-2mm}\mathrm{C2}:\,\, \frac{\Tr(\mathbf{W}\mathbf{H})}
{\Gamma_{\mathrm{req}} } \geq \Tr(\mathbf{V}\mathbf{H})+\sigma_{\mathrm{s}}^2 ,\notag\\
\hspace*{-1mm}&&\hspace*{-2mm}\mathrm{C3, C4},\notag\\
\hspace*{-1mm}&&\hspace*{-2mm}\mathrm{C5}:\,\, \Tr\Big((\mathbf{W}+\mathbf{V})\mathbf{G}_j\mathbf{G}_j^H\Big) \geq \tau_j,\notag \forall j,\\
\hspace*{-5mm}&&\hspace*{-2mm}\mathrm{C6}:\,\, \Rank(\mathbf{W})=1,\,\, \mathrm{C7}:\,\, \mathbf{W}\succeq\zero,
\end{eqnarray}\vspace*{-3mm}
\end{Prob}\noindent
where $\mathbf{W}=\mathbf{w}\mathbf{w}^H$ is a new optimization variable matrix  and $\bm\tau=[\tau_1,\tau_2,\ldots,\tau_J]$ is a vector of auxiliary optimization variables. The extra constraint C5 represents the hypograph of the inner loop optimization problem.
\begin{table}[t]\caption{Iterative Resource Allocation Algorithm.}\label{table:algorithm}
\begin{algorithm} [H]                    
\renewcommand\thealgorithm{}
\caption{}          

\label{alg1}                           
\begin{algorithmic} [1]
\STATE Initialize the maximum number of iterations $L_{\max}$, iteration index $n=0$, $\bm\mu$, and $\bm \beta$

\REPEAT [Outer Loop]
\STATE Solve the inner loop problem in \eqref{eqn:SDP_relaxed_problem} via semidefinite program relaxation for
 given $(\bm\mu^n,\bm\beta^n)$ and obtain the intermediate beamformer $\mathbf{w}'$ and artificial noise covariance matrix $\mathbf{V}'$
\IF { \eqref{eqn:convergence_condition} is satisfied} \RETURN
Optimal beamformer $\mathbf{w}^*=\mathbf{w}'$ and artificial noise covariance matrix $\mathbf{V}^*=\mathbf{V}'$
 \ELSE \STATE
Update $\bm \mu$ and $\bm\beta$ according to \eqref{eqn:update_beta} and $n=n+1$
 \ENDIF
 \UNTIL{\eqref{eqn:conditions1} and \eqref{eqn:conditions2} are satisfied $\,$or $n=L_{\max}$}

\end{algorithmic}
\end{algorithm}
\end{table}

We note that the inner loop problem in  \eqref{eqn:TP_maximization-transfomred} is still a non-convex optimization problem. In particular, the non-convexity arises from the log-det function in C3 and the combinatorial rank constraint C6. To circumvent
the non-convexity, we first introduce the following proposition to handle constraint C3.

\begin{proposition}\label{prop:relaxed_c3} For $R_{\mathrm{ER}}^{\mathrm{Tol}}> 0,\forall j$, and $\Rank(\mathbf{W})\le 1$, constraint $\mathrm{C3}$ is equivalent to constraint $\overline{\mathrm{C3}}$, i.e.,
\begin{eqnarray}\label{eqn:det_to_matrix}
\mathrm{C3}\Leftrightarrow\overline{\mathrm{C3}}\mbox{: } \mathbf{G}_j^H\mathbf{W}\mathbf{G}_j&\preceq& \alpha_{\mathrm{ER}} \mathbf{Q}_{j},\,\, \forall j,
\end{eqnarray}where $\alpha_{\mathrm{ER}}=2^{R_{\mathrm{ER}}^{\mathrm{Tol}}}-1$ is an auxiliary constant and $\overline{\mbox{C3}}$
 is a linear matrix inequality (LMI) constraint.
\end{proposition}

\,\,\emph{Proof:} Please refer to Appendix A in \cite{JR:Kwan_CR_Layered} for the proof.\qed

Now, we apply Proposition \ref{prop:relaxed_c3} to Problem \eqref{eqn:TP_maximization-transfomred} by replacing constraint $\mbox{C3}$ with constraint $\overline{\mbox{C3}}$ which yields:
\begin{Prob}{\normalfont Equivalent Formulation of Problem \eqref{eqn:TP_maximization-transfomred}}\label{Prob:equivalent_max_min}
\begin{eqnarray}\notag\label{eqn:equivalent}
\underset{\mathbf{W},\mathbf{V}\in \mathbb{H}^{N_{\mathrm{T}}},\bm\tau}{\maxo}\,\, &&\sum_{j=1}^J \mu_j^*\Big[\hspace*{-0.5mm}M_j\hspace*{-0.5mm}- \hspace*{-0.5mm} \beta_j^*\Big(1+\exp\big(\hspace*{-0.5mm}-\hspace*{-0.5mm}a_j(\tau_j\hspace*{-0.5mm}-\hspace*{-0.5mm}b_j)\big)\Big)\hspace*{-0.5mm}\Big]\\
\hspace*{-1mm}\mathrm{subject\,\,to}\,\, &&\hspace*{-2mm}\mathrm{C1, C2, C4, C5, C7},\notag\\
\hspace*{-1mm}&&\hspace*{-2mm}\overline{\mathrm{C3}}: \mathbf{G}_j^H\mathbf{W}\mathbf{G}_j\preceq \alpha_{\mathrm{ER}} \mathbf{Q}_{j},\,\, \forall j,\notag\\
\hspace*{-5mm}&&\hspace*{-2mm}\mathrm{C6}:\,\, \Rank(\mathbf{W})=1.
\end{eqnarray}
\end{Prob}
The non-convexity of  \eqref{eqn:equivalent} is now only due to the rank constraint in C6. We adopt semidefinite programming (SDP) relaxation  to obtain a tractable solution. Specifically, we remove the non-convex constraint C6 from \eqref{eqn:equivalent} which yields:

\begin{Prob}{\normalfont SDP Relaxation of Problem \eqref{eqn:equivalent}}\label{Prob:equivalent_max_min}
\begin{eqnarray}\notag\label{eqn:SDP_relaxed_problem}
\underset{\mathbf{W},\mathbf{V}\in \mathbb{H}^{N_{\mathrm{T}}},\bm\tau}{\maxo}\,\, &&\sum_{j=1}^J \mu_j^*\Big[\hspace*{-0.5mm}M_j\hspace*{-0.5mm}- \hspace*{-0.5mm} \beta_j^*\Big(1+\exp\big(\hspace*{-0.5mm}-\hspace*{-0.5mm}a_j(\tau_j\hspace*{-0.5mm}-\hspace*{-0.5mm}b_j)\big)\Big)\hspace*{-0.5mm}\Big]\\
\hspace*{-1mm}\mathrm{subject\,\,to}\,\, &&\hspace*{-2mm}\mathrm{C1, C2, C4, C5, C7},\notag\\
\hspace*{-1mm}&&\hspace*{-2mm}\overline{\mathrm{C3}}: \mathbf{G}_j^H\mathbf{W}\mathbf{G}_j\preceq \alpha_{\mathrm{ER}} \mathbf{Q}_{j},\,\, \forall j,\notag\\
\hspace*{-5mm}&&\hspace*{-2mm}\cancel{\mathrm{C6}:\,\, \Rank(\mathbf{W})=1}.
\end{eqnarray}
\end{Prob}
In fact, \eqref{eqn:SDP_relaxed_problem} is a standard convex optimization problem which can be solved by numerical convex program solvers such as Sedumi or SDPT3 \cite{website:CVX}. Now, we study the tightness of the adopted SDP relaxation in \eqref{eqn:SDP_relaxed_problem}.
\begin{Thm}\label{thm:rankone}
For $\Gamma_{\mathrm{req}}>0$ and if the considered problem is feasible,  we can construct a rank-one solution of \eqref{eqn:equivalent} based on the solution of \eqref{eqn:SDP_relaxed_problem}.
\end{Thm}

\,\,\emph{Proof:} Please refer to the Appendix. \qed

Therefore, the non-convex optimization problem in \eqref{eqn:TP_maximization-transfomred} can be solved optimally.

\subsection{Solution of the Outer Loop Problem}
In this section, an iterative algorithm based on the damped Newton method is adopted  to  update $(\bm{\mu},\bm{\beta})$  for the outer loop problem. For notational simplicity, we define functions $\varphi_j(\beta_j)=\beta_j\Big(1+\exp\big(-a_j(P_{\mathrm{ER}_j}-b_j)\big)\Big)-M_j$
and $\varphi_{J+i}(\mu_i)=\mu_i\Big(1+\exp\big(-a_i(P_{\mathrm{ER}_i}-b_i)\big)\Big)-1$, $i\in\{1,\ldots,J\}$. It is shown in \cite{JR:sum_of_ratios} that the unique optimal solution  $(\bm{\mu}^*,\bm{\beta}^*)$ is obtained if and only if  $\bm\varphi(\bm \mu,  \bm\beta)=[\varphi_1,\varphi_2,\ldots,\varphi_{2J}]^T=\zero$.  Therefore, in the $n$-th iteration of the iterative algorithm, ${\bm \mu}^{n+1}$ and ${\bm \beta}^{n+1}$ can be updated as, respectively,
\begin{eqnarray}\notag
{\bm \mu}^{n+1}\hspace*{-1.0mm}&=&\hspace*{-1.0mm}{\bm \mu}^{n}+\zeta^n\mathbf{q}^n_{J+1:2J}\quad\mbox{and}\quad  {\bm \beta}^{n+1}={\bm \beta}^{n}+\zeta^n\mathbf{q}^n_{1:J},\\
\mbox{where }\,\,\mathbf{q}^n\hspace*{-1.0mm}&=&\hspace*{-1.0mm}[\bm\varphi'(\bm{\mu},\bm{\beta})]^{-1}\bm\varphi(\bm{\mu},\bm{\beta})\label{eqn:update_beta}
\end{eqnarray}
 and $\bm\varphi'(\bm{\mu},\bm{\beta})$ is the Jacobian matrix of $\bm\varphi(\bm{\mu},\bm{\beta})$. $\zeta^n$ is the largest  $\varepsilon^l$ satisfying
\begin{eqnarray}\label{eqn:convergence_condition}
&&\norm{\bm\varphi\big({\bm \mu}^{n}+\varepsilon^l\mathbf{q}^n_{J+1:2J},{\bm \beta}^{n}+\varepsilon^l\mathbf{q}^n_{1:J}\big)}\notag\\
&&\leq (1-\eta\varepsilon^l)\norm{\bm\varphi(\bm{\mu},\bm{\beta})},
\end{eqnarray}
where $l\in\{1,2,\ldots\}$, $\varepsilon^l\in(0,1)$, and $\eta\in(0,1)$. The damped Newton method converges to the unique solution $(\bm{\mu}^*,\bm{\beta}^*)$  satisfying the system of equations \eqref{eqn:conditions1} and \eqref{eqn:conditions2}, cf. \cite{JR:sum_of_ratios}.

\section{Results}\label{sect:simulation}
In this section, simulation results are presented to illustrate the performance of the proposed
resource allocation algorithm. We summarize the most important simulation parameters in Table \ref{table:parameters}.
In the simulation,  the IR  and the $J=10$ ERs  are located at $50$ meters and $10$ meters from the BS, respectively. The maximum tolerable data rate at the potential eavesdropper is set to $R_{\mathrm{ER}}^{\mathrm{Tol}}=1$ bit/s/Hz. For the non-linear EH circuit, we set $M_j=20$ mW which corresponds to the maximum harvested power per ER. Besides, we adopt $a_j=6400$ and $b_j=0.003$.

\begin{table}[t]
\caption{Simulation Parameters} \label{table:parameters}
\centering
\setlength\tabcolsep{2pt}
\begin{tabular}{ | l | l | } \hline

      System bandwidth                                          & $200$ kHz \\ \hline 
            Carrier center frequency                           & $
      915$ MHz\\ \hline
      Transceiver  antenna gain                                     & $10$ dBi \\ \hline
      Number of receive antennas  $N_\mathrm{R}$                                  & $2$ \\ \hline
      Noise power              $\sigma^2$                          & $ -95$ dBm \\ \hline
          Maximum transmit power    $P_{\max}$                                  & $36$ dBm \\ \hline
      \pbox{20cm}{ BS-to-ER \\ fading distribution }                                     & Rician with Rician factor $3$ dB \\
        \hline
\end{tabular}
\end{table}

\begin{figure}[t]
        \centering
       \includegraphics[width=3.5 in]{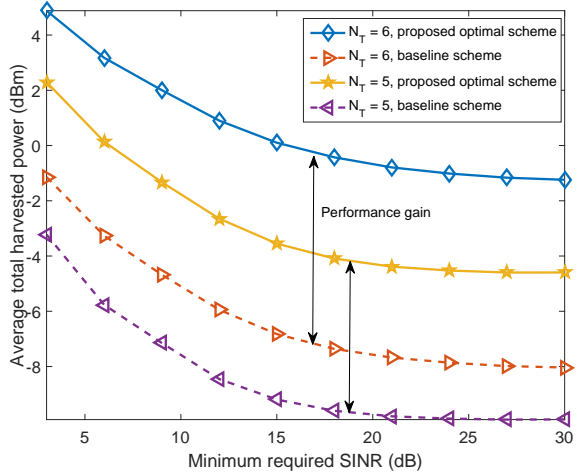}
        \caption{Average total harvested power (dBm) versus the minimum required SINR (dB).}\label{fig:HP_SINR}
        \end{figure}
In Figure \ref{fig:HP_SINR}, we study the average total harvested power versus the minimum required receive SINR, $\Gamma_{\mathrm{req}}$, at the IR  for different numbers of transmit antennas  and resource allocation schemes. As can be observed, the average total harvested power decreases with increasing $\Gamma_{\mathrm{req}}$. Indeed, to satisfy a more stringent SINR requirement,  the direction of
information beam has to be steered towards the IR which yields a smaller amount of RF energy for
EH at the ERs. On the other hand,    a significant EH gain can be achieved by the proposed optimal scheme when the number of antennas equipped at the  BS  increases. In fact, additional transmit antennas equipped at the BS provide extra spatial degrees of freedom which facilitate a more flexible resource allocation. In particular,  the  BS  can steer the direction of the artificial noise and the information signal towards the ERs  accurately to improve the WET  efficiency.
For comparison, we also show  the performance of a baseline scheme. For the baseline scheme, the resource allocation algorithm is designed based on an existing linear EH model, cf. \eqref{eqn:linear_EH_model}. Specifically, we optimize $\mathbf{w},\mathbf{V}$ to maximize the total harvested power    subject to the constraints in \eqref{eqn:TP_maximization}. Then, this baseline scheme is applied for resource allocation in the considered system with non-linear ERs.  We observe from Figure \ref{fig:HP_SINR} that
 a substantial performance gain is achieved by the proposed optimal resource allocation algorithm compared to the baseline scheme. This is due to the fact that   resource allocation mismatch occurs in the baseline scheme  as it does not account for the non-linear nature of the EH circuits.

        \begin{figure}[t]        \centering
        \includegraphics[width=3.5 in]{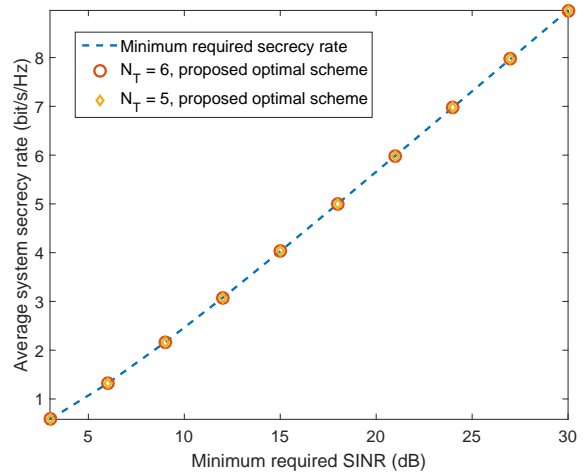}
        \caption{Average system secrecy rate (bit/s/Hz) versus the minimum required SINR (dB).}
        \label{fig:rate_SINR}
\end{figure}

Figure \ref{fig:rate_SINR} illustrates the average system secrecy rate versus the minimum required SINR $\Gamma_{\mathrm{req}}$ of the IR for different numbers of transmit antennas, $N_{\mathrm{T}}$.  The average system secrecy rate, i.e., $C_{\mathrm{sec}}$, increases with  increasing $\Gamma_{\mathrm{req}}$. This is because  the maximum achievable rate of the ERs for decoding the IR signal is limited by the resource allocation to be less than $R_{\mathrm{ER}}^{\mathrm{Tol}}=1$ bit/s/Hz. Besides, although the  minimum SINR requirement increases in Figure \ref{fig:rate_SINR},  the proposed optimal scheme is able to fulfill all QoS requirements due to the proposed optimization framework.

\section{Conclusions}\label{sect:conclusion}
In this paper, a resource allocation algorithm enabling secure SWIPT in IoT communication networks was presented.  The algorithm design based on a practical non-linear EH model was formulated as a non-convex optimization problem for the maximization of the total energy transferred to the ERs. We transformed the resulting non-convex optimization
problem  into  two  nested   optimization  problems  which  led  to  an  efficient  iterative approach
for  obtaining  the globally optimal  solution. Numerical
results unveiled the potential performance gain in EH   brought  by  the
proposed optimization and its robustness against
eavesdropping for IoT applications.


\section*{Appendix-Proof of Theorem \ref{thm:rankone}}\label{app:rankone}
To start with, we first define $\bm\tau^*$ as the optimal objective value  of \eqref{eqn:SDP_relaxed_problem}.
When $\Rank(\mathbf{W})>1$ holds after solving \eqref{eqn:SDP_relaxed_problem},  an optimal rank-one solution for \eqref{eqn:SDP_relaxed_problem}  can  be constructed as follows \cite{7463025}. For a given  $\bm\tau^*$,  we solve an auxiliary optimization problem:
\begin{eqnarray} \label{eqn:dummy_problem}
\underset{{\mathbf{W},\mathbf{V}\in \mathbb{H}^{N_{\mathrm{T}}}}}{\mino}\,\, && \Tr(\mathbf{W})\\
\mathrm{subject\,\,to}\,\, &&\hspace*{-2mm}\mathrm{C1, C2, \overline{\mathrm{C3}}, C4, C5, C7}.\notag
\end{eqnarray}
It can be observed  that the optimal solution of \eqref{eqn:dummy_problem} is also an optimal  resource optimal resource allocation policy for \eqref{eqn:SDP_relaxed_problem} when $\bm\tau^*$ is fixed in \eqref{eqn:dummy_problem}. Therefore, in the remaining part of the proof,  we show that solving \eqref{eqn:dummy_problem} returns a rank-one beamforming matrix $\mathbf{W}$. Therefore, we study the Lagrangian of problem \eqref{eqn:dummy_problem} which is given by:
\begin{eqnarray}
L&=&\Tr(\mathbf{W})+\lambda(\Tr(\mathbf{W}+\mathbf{V})- P_{\mathrm{max}}) - \Tr(\mathbf{W}\mathbf{R})\notag\\
&-&\sum_{j=1}^J\rho_j \Bigg(\tau_j^*-\Tr\Big((\mathbf{W}+\mathbf{V})\mathbf{G}_j\mathbf{G}_j^H\Big)\Bigg)- \Tr(\mathbf{V}\mathbf{Z})\notag\\
&+&\sum_{j=1}^J\Tr(\mathbf{D}_{\mathrm{C}_{3_j}}( \mathbf{G}_j^H\mathbf{W}\mathbf{G}_j-\alpha_{\mathrm{ER}} \mathbf{Q}_{j}))\notag\\
&+&\alpha\Big( \Tr(\mathbf{V}\mathbf{H})+\sigma_{\mathrm{s}}^2-\frac{\Tr(\mathbf{W}\mathbf{H})}
{\Gamma_{\mathrm{req}} }\Big)+{\bm\Delta},
\end{eqnarray}
where $\lambda\geq 0$, $\alpha\geq 0$, $\mathbf{D}_{\mathrm{C}_{3_j}}\succeq\zero,\forall j\in\{1,\ldots,J\}$, $\mathbf{Z}\succeq\zero$, $\rho_j\geq 0$, and $\mathbf{R}\succeq\zero$ are the dual variables for constraints C1, C2, $\overline{\mathrm{C3}}$, C4, C5, and C7, respectively.  $\bm\Delta$ is a collection of variables and constants that are not relevant to the proof.

Then, we exploit the following  Karush-Kuhn-Tucker (KKT) conditions which are needed for the proof\footnote{In this proof, the optimal
primal and dual variables of \eqref{eqn:dummy_problem} are denoted by the corresponding
variables with an asterisk superscript.}:

\begin{eqnarray}
\hspace*{-8mm}&&\mathbf{R}^*,\mathbf{Z}^*,\mathbf{D}_{\mathrm{C}_{3_j}}^*\succeq \zero,\quad\lambda^*,\alpha,\rho^*_j\ge0,\\
\hspace*{-8mm}&& \mathbf{R^*W^*}=\zero,\quad \label{eqn:KKT-complementarity}  \mathbf{Z^*V^*}=\zero, \label{eqn:KKT-complementarity}\\
\label{eqn:KKT_Y1}
\hspace*{-8mm}&&\mathbf{R^*}=
(\lambda^*+1)\mathbf{I}_{N_\mathrm{T}}-\sum_{j=1}^J\rho_j\mathbf{G}_j\mathbf{G}_j^H +\sum_{j=1}^J \mathbf{G}_j\mathbf{D}_{\mathrm{C}_{3_j}}^*  \mathbf{G}_j^H\notag\\
\hspace*{-8mm}&&\hspace*{5mm}-\alpha^*\frac{\mathbf{H}}
{\Gamma_{\mathrm{req}} },\\
\label{eqn:KKT_Y2}\hspace*{-8mm}&&\mathbf{Z^*}= \lambda^*\mathbf{I}_{N_\mathrm{T}}-\sum_{j=1}^J\rho_j\mathbf{G}_j\mathbf{G}_j^H \notag\\
\hspace*{-8mm}&&\hspace*{5mm}-\sum_{j=1}^J \alpha_{\mathrm{ER}}\mathbf{G}_j\mathbf{D}_{\mathrm{C}_{3_j}}^*  \mathbf{G}_j^H+\alpha^* \mathbf{H}.
\end{eqnarray}

Then, subtracting  \eqref{eqn:KKT_Y2} from \eqref{eqn:KKT_Y1} yields
\begin{eqnarray}\label{eqn:temp_eq}
\mathbf{R}^*&=&\underbrace{\mathbf{I}_{N_\mathrm{T}}+\mathbf{Z}^*+ \sum_{j=1}^J(1+\alpha_{\mathrm{ER}}) \mathbf{G}_j\mathbf{D}_{\mathrm{C}_{3_j}}^*  \mathbf{G}_j^H}_{\mathbf{A}\succ \zero}\notag\\
&-& \alpha^*\mathbf{H}\Big(1+\frac{1}
{\Gamma_{\mathrm{req}} }\Big).
\end{eqnarray}
Besides,  constraint C2 in \eqref{eqn:dummy_problem} is satisfied with equality for the optimal solution and we have $\alpha^*>0$. From \eqref{eqn:temp_eq}, we have
\begin{eqnarray}\notag
&&\Rank(\mathbf{R}^*)+\Rank(\alpha^*\mathbf{H}(1+1/
{\Gamma_{\mathrm{req}} }))\\
 \notag&\ge &\Rank(\mathbf{R}^*+\alpha^*\mathbf{H}(1+1/
{\Gamma_{\mathrm{req}} }))\\
&= &\Rank(\mathbf{A})=N_\mathrm{T}\notag\\
&\Rightarrow& \Rank(\mathbf{R}^*)\ge N_\mathrm{T}-1.
\end{eqnarray}
As a result,  $\Rank(\mathbf{R}^*)$ is either  $N_\mathrm{T}-1$ or $N_\mathrm{T}$. Furthermore, since $\Gamma_{\mathrm{req}}>0$ in C2,  $\mathbf{W}^*\ne\mathbf{0}$ is necessary. Hence, $\Rank(\mathbf{R}^*)=N_{\mathrm{T}}-1$ and $\Rank(\mathbf{W}^*)=1$ hold and the rank-one solution for \eqref{eqn:SDP_relaxed_problem} is constructed.  \qed



\begin{thebibliography}{}
\providecommand{\url}[1]{#1}
\csname url@samestyle\endcsname
\providecommand{\newblock}{\relax}
\providecommand{\bibinfo}[2]{#2}
\providecommand{\BIBentrySTDinterwordspacing}{\spaceskip=0pt\relax}
\providecommand{\BIBentryALTinterwordstretchfactor}{4}
\providecommand{\BIBentryALTinterwordspacing}{\spaceskip=\fontdimen2\font plus
\BIBentryALTinterwordstretchfactor\fontdimen3\font minus
  \fontdimen4\font\relax}
\providecommand{\BIBforeignlanguage}[2]{{%
\expandafter\ifx\csname l@#1\endcsname\relax
\typeout{** WARNING: IEEEtran.bst: No hyphenation pattern has been}%
\typeout{** loaded for the language `#1'. Using the pattern for}%
\typeout{** the default language instead.}%
\else
\language=\csname l@#1\endcsname
\fi
#2}}
\providecommand{\BIBdecl}{\relax}
\BIBdecl

\end{thebibliography}


\begin{thebibliography}{10}
\providecommand{\url}[1]{#1}
\csname url@samestyle\endcsname
\providecommand{\newblock}{\relax}
\providecommand{\bibinfo}[2]{#2}
\providecommand{\BIBentrySTDinterwordspacing}{\spaceskip=0pt\relax}
\providecommand{\BIBentryALTinterwordstretchfactor}{4}
\providecommand{\BIBentryALTinterwordspacing}{\spaceskip=\fontdimen2\font plus
\BIBentryALTinterwordstretchfactor\fontdimen3\font minus
  \fontdimen4\font\relax}
\providecommand{\BIBforeignlanguage}[2]{{%
\expandafter\ifx\csname l@#1\endcsname\relax
\typeout{** WARNING: IEEEtran.bst: No hyphenation pattern has been}%
\typeout{** loaded for the language `#1'. Using the pattern for}%
\typeout{** the default language instead.}%
\else
\language=\csname l@#1\endcsname
\fi
#2}}
\providecommand{\BIBdecl}{\relax}
\BIBdecl

\bibitem{JR:IOT}
M.~Zorzi, A.~Gluhak, S.~Lange, and A.~Bassi, ``{From today's INTRAnet of things
  to a Future INTERnet of Things: a Wireless- and Mobility-Related View},''
  \emph{{IEEE Wireless Commun.}}, vol.~17, pp. 44--51, 2010.

\bibitem{JR:Kwan_hybrid_BS}
D.~W.~K. Ng, E.~S. Lo, and R.~Schober, ``{Energy-Efficient Resource Allocation
  in OFDMA Systems with Hybrid Energy Harvesting Base Station},'' \emph{IEEE
  Trans. Wireless Commun.}, vol.~12, pp. 3412--3427, Jul. 2013.

\bibitem{CN:Shannon_meets_tesla}
P.~Grover and A.~Sahai, ``{Shannon Meets Tesla: Wireless Information and Power
  Transfer},'' in \emph{Proc. IEEE Intern. Sympos. on Inf. Theory}, Jun. 2010,
  pp. 2363 --2367.

\bibitem{Krikidis2014}
I.~Krikidis, S.~Timotheou, S.~Nikolaou, G.~Zheng, D.~W.~K. Ng, and R.~Schober,
  ``{Simultaneous Wireless Information and Power Transfer in Modern
  Communication Systems},'' \emph{IEEE Commun. Mag.}, vol.~52, no.~11, pp.
  104--110, Nov. 2014.

\bibitem{Ding2014}
Z.~Ding, C.~Zhong, D.~W.~K. Ng, M.~Peng, H.~A. Suraweera, R.~Schober, and H.~V.
  Poor, ``{Application of Smart Antenna Technologies in Simultaneous Wireless
  Information and Power Transfer},'' \emph{IEEE Commun. Mag.}, vol.~53, no.~4,
  pp. 86--93, Apr. 2015.

\bibitem{JR:SWIPT_mag}
X.~Chen, Z.~Zhang, H.-H. Chen, and H.~Zhang, ``{Enhancing Wireless Information
  and Power Transfer by Exploiting Multi-Antenna Techniques},'' \emph{IEEE
  Commun. Mag.}, no.~4, pp. 133--141, Apr. 2015.

\bibitem{JR:SWIPT_mag_Ming_Kwan}
X.~Chen, D.~W.~K. Ng, and H.-H. Chen, ``{Secrecy Wireless Information and Power
  Transfer: Challenges and Opportunities},'' \emph{IEEE Commun. Mag.}, 2016.

\bibitem{JR:QQ_WPC}
Q.~Wu, M.~Tao, D.~Ng, W.~Chen, and R.~Schober, ``{Energy-Efficient Resource
  Allocation for Wireless Powered Communication Networks},'' \emph{IEEE Trans.
  Wireless Commun.}, vol.~15, pp. 2312--2327, Mar. 2016.

\bibitem{COML:EE_WIPT}
X.~Chen, X.~Wang, and X.~Chen, ``{Energy-Efficient Optimization for Wireless
  Information and Power Transfer in Large-Scale MIMO Systems Employing Energy
  Beamforming},'' \emph{IEEE Wireless Commun. Lett.}, vol.~2, pp. 1--4, Dec.
  2013.

\bibitem{JR:MIMO_WIPT}
R.~Zhang and C.~K. Ho, ``{MIMO Broadcasting for Simultaneous Wireless
  Information and Power Transfer},'' \emph{IEEE Trans. Wireless Commun.},
  vol.~12, pp. 1989--2001, May 2013.

\bibitem{JR:QQ_WPCN}
Q.~Wu, W.~Chen, and J.~Li, ``{Wireless Powered Communications With Initial
  Energy: QoS Guaranteed Energy-Efficient Resource Allocation},'' \emph{IEEE
  Wireless Commun. Lett.}, vol.~19, Dec. 2015.

\bibitem{Report:Wire_tap}
A.~D. Wyner, ``{The Wire-Tap Channel},'' Tech. Rep., Oct. 1975.

\bibitem{JR:Massive_MIMO}
J.~Zhu, R.~Schober, and V.~Bhargava, ``{Secure Transmission in Multicell
  Massive MIMO Systems},'' \emph{IEEE Trans. Wireless Commun.}, vol.~13, pp.
  4766--4781, Sep. 2014.

\bibitem{JR:Artifical_Noise1}
S.~Goel and R.~Negi, ``{Guaranteeing Secrecy using Artificial Noise},''
  \emph{IEEE Trans. Wireless Commun.}, vol.~7, pp. 2180 -- 2189, Jun. 2008.

\bibitem{JR:HM_security_1}
H.~M. Wang, C.~Wang, D.~Ng, M.~Lee, and J.~Xiao, ``{Artificial Noise Assisted
  Secure Transmission for Distributed Antenna Systems},'' \emph{IEEE Trans.
  Signal Process.}, vol.~PP, no.~99, pp. 1--1, 2016.

\bibitem{JR:HM_security_2}
J.~Chen, X.~Chen, W.~H. Gerstacker, and D.~W.~K. Ng, ``{Resource Allocation for
  a Massive MIMO Relay Aided Secure Communication},'' \emph{{IEEE Trans. on
  Inf. Forensics and Security}}, vol.~11, no.~8, pp. 1700--1711, Aug 2016.

\bibitem{JR:Kwan_secure_imperfect}
D.~W.~K. Ng, E.~S. Lo, and R.~Schober, ``{Robust Beamforming for Secure
  Communication in Systems with Wireless Information and Power Transfer},''
  \emph{IEEE Trans. Wireless Commun.}, vol.~13, pp. 4599--4615, Aug. 2014.

\bibitem{JR:MOOP_SWIPT}
------, ``{Multiobjective Resource Allocation for Secure Communication in
  Cognitive Radio Networks With Wireless Information and Power Transfer},''
  \emph{IEEE Trans. Veh. Technol.}, vol.~65, no.~5, pp. 3166--3184, May 2016.

\bibitem{CN:EH_measurement_2}
J.~Guo and X.~Zhu, ``{An Improved Analytical Model for RF-DC Conversion
  Efficiency in Microwave Rectifiers},'' in \emph{IEEE MTT-S Int. Microw. Symp.
  Dig.}, Jun. 2012, pp. 1--3.

\bibitem{JR:Energy_harvesting_circuit}
C.~Valenta and G.~Durgin, ``{Harvesting Wireless Power: Survey of
  Energy-Harvester Conversion Efficiency in Far-Field, Wireless Power Transfer
  Systems},'' \emph{IEEE Microw. Mag.}, vol.~15, pp. 108--120, Jun. 2014.

\bibitem{JR:EH_measurement_1}
T.~Le, K.~Mayaram, and T.~Fiez, ``{Efficient Far-Field Radio Frequency Energy
  Harvesting for Passively Powered Sensor Networks},'' \emph{IEEE J.
  Solid-State Circuits}, vol.~43, pp. 1287--1302, May 2008.

\bibitem{JR:non_linear_model}
E.~Boshkovska, D.~Ng, N.~Zlatanov, and R.~Schober, ``{Practical Non-Linear
  Energy Harvesting Model and Resource Allocation for SWIPT Systems},''
  \emph{IEEE Commun. Lett.}, vol.~19, pp. 2082--2085, Dec. 2015.

\bibitem{JR:Elena_TCOM}
\BIBentryALTinterwordspacing
E.~Boshkovska, D.~W.~K. Ng, N.~Zlatanov, A.~Koelpin, and R.~Schober, ``{Robust
  Resource Allocation for {MIMO} Wireless Powered Communication Networks Based
  on a Non-linear {EH} Model},'' 2016, submitted to TCOM. [Online]. Available:
  \url{http://arxiv.org/abs/1609.03836}
\BIBentrySTDinterwordspacing

\bibitem{JR:sum_of_ratios}
\BIBentryALTinterwordspacing
Y.~Jong, ``{An Efficient Global Optimization Algorithm for Nonlinear
  Sum-of-Ratios Problem},'' May 2012. [Online]. Available:
  \url{http://www.optimization-online.org/DB FILE/2012/08/3586.pdf}
\BIBentrySTDinterwordspacing

\bibitem{JR:Kwan_CR_Layered}
D.~W.~K. Ng, M.~Shaqfeh, R.~Schober, and H.~Alnuweiri, ``{Robust Layered
  Transmission in Secure MISO Multiuser Unicast Cognitive Radio Systems},''
  \emph{IEEE Trans. Veh. Technol.}, vol.~65, no.~10, pp. 8267--8282, Oct. 2016.

\bibitem{website:CVX}
M.~Grant and S.~Boyd, ``{CVX: Matlab Software for Disciplined Convex
  Programming, version 2.0 Beta},'' [Online] \url{https://cvxr.com/cvx}, Sep.
  2013.

\bibitem{7463025}
Y.~Sun, D.~W.~K. Ng, J.~Zhu, and R.~Schober, ``{Multi-Objective Optimization
  for Robust Power Efficient and Secure Full-Duplex Wireless Communication
  Systems},'' \emph{IEEE Trans. Wireless Commun.}, vol.~15, no.~8, pp.
  5511--5526, Aug. 2016.

\end{thebibliography}
\end{document}